\documentclass[twocolumn,aps,pra,superscriptaddress,showpacs,tightenlines]{revtex4-1}
\usepackage{mathptmx}
\DeclareMathAlphabet{\mathcal}{OMS}{cmsy}{m}{n}
\usepackage{mathrsfs}

\usepackage[T1]{fontenc}
\usepackage[latin9]{inputenc}
\usepackage{amstext}
\usepackage{amsmath}
\usepackage{graphicx}
\usepackage{esint}
\usepackage{color}
\usepackage[colorlinks, linkcolor=blue, citecolor=blue,hyperindex,bookmarks=false,pdfstartview=FitH]{hyperref}
\makeatletter

\@ifundefined{textcolor}{}
{%
 \definecolor{BLACK}{gray}{0}
\definecolor{WHITE}{gray}{1}
 \definecolor{RED}{rgb}{1,0,0}
 \definecolor{GREEN}{rgb}{0,1,0}
 \definecolor{BLUE}{rgb}{0,0,1}
 \definecolor{CYAN}{cmyk}{1,0,0,0}
 \definecolor{MAGENTA}{cmyk}{0,1,0,0}
 \definecolor{YELLOW}{cmyk}{0,0,1,0}
}
\makeatother

\begin{document}

\title{Switchable bipartite and genuine tripartite entanglement via an optoelectromechanical interface}
\author{Cheng Jiang}
\affiliation{School of Physics and Electronic Electrical Engineering, Huaiyin Normal University, 111 West Chang Jiang Road, Huai'an 223300, China}
\affiliation{Beijing Computational Science Research Center, Beijing 100193, China}

\author{Spyros Tserkis}
\affiliation{Centre for Quantum Computation and Communication Technology, School of Mathematics and Physics, University of Queensland, St. Lucia, Queensland 4072, Australia}

\author{Kevin Collins}
\affiliation{School of Natural Sciences, University of California, Merced, 5200 North Lake Road, Merced,
California 95343, USA}

\author{Sho Onoe}
\affiliation{Centre for Quantum Computation and Communication Technology, School of Mathematics and Physics, University of Queensland, St. Lucia, Queensland 4072, Australia}

\author{Yong Li}
\affiliation{Beijing Computational Science Research Center, Beijing 100193, China}

\author{Lin Tian}
\email{ltian@ucmerced.edu}
\affiliation{School of Nature Sciences, University of California, Merced, California 95343, USA}

\begin{abstract}
Controllable multipartite entanglement is a crucial element in quantum information processing. Here we present a scheme that generates switchable bipartite and genuine tripartite entanglement between microwave and optical photons via an optoelectromechanical interface, where microwave and optical cavities are coupled to a mechanical mode with controllable coupling constants.
We show that by tuning an effective gauge phase between the coupling constants to the ``sweet spots'', bipartite entanglement can be generated and switched between designated output photons. The bipartite entanglement is robust against the mechanical noise and the signal loss to the mechanical mode when the couplings are chosen to satisfy the impedance matching condition.
When the gauge phase is tuned away from the ``sweet spots'', genuine tripartite entanglement can be generated and verified with homodyne measurement on the quadratures of the output fields. Our result can lead to the implementation of controllable and robust multipartite entanglement in hybrid quantum systems operated in distinctively different frequencies.
\end{abstract}
\maketitle

\section{Introduction}
Entanglement is a profound feature in quantum theory and an indispensable resource in quantum communication and quantum networks~\cite{Horodecki, Kimble}. Bipartite and tripartite entanglement for qubits and continuous variables has been demonstrated in a variety of physical systems, such as superconducting qubits~\cite{Berkley, Steffen}, atomic ensembles~\cite{Julsgaard}, and optical modes  ~\cite{ArmstrongPKLamNatPhys2015, ShalmJenneweinNatPhys2013}. The generation of controllable entanglement between systems of distinctively different frequencies such as microwave and optical photons is crucial for hybrid quantum computing, where different systems are bridged together to boost the overall performance of the quantum devices. However, due to the difficulty of interfacing different systems in a noiseless and lossless manner, it is often challenging to generate switchable and robust entanglement in hybrid quantum systems. For example, when coupling a superconducting qubit to an optical device, extreme care has to be taken to prevent stray photons from exciting quasiparticles and destroying the quantum coherence of the qubit~\cite{Tian2004}.

Opto- and electro-mechanical systems provide an excellent candidate for an interface that connects different components in hybrid quantum systems~\cite{Stannigel2010, Metcalfe2014, Tian2015, WangPRL2012, TianPRL2012, BarzanjehPRL2012}. Because of the ubiquitous existence of mechanical vibrations, mechanical resonators can be coupled to devices of a broad spectral range from acoustic to infrared frequencies via radiation pressure force~\cite{AspelmeyerRMP}.
Furthermore, the optical (electrical) response of an opto- (electro-) mechanical system can be tuned by applying strong driving fields in selected phonon sidebands to the cavity mode, which gives us a toolbox to manipulate the quantum states of the cavity and the mechanical modes.
Under red-detuned driving, the phenomena of optomechanically-induced transparency~\cite{Weis,Naeini1}, sideband cooling~\cite{Chan,Teufel}, and quantum state conversion~\cite{Andrews2014, Cleland2013, Polzik2014} have been demonstrated. With quantum-engineered time-reversal symmetry to generate uni-directionality, nonreciprocal transmission via opto- and electro-mechanical interface has been studied recently~\cite{Hafezi, ShenZ, Ruesink, Miri, TianPRA2017, XuXW2, Bernier, Barzanjeh2017, Teufel2017,  FangKJ, Malz, Sillanpaa2018, LiY, ZhangXZ, JiangOE, ShenZ2, Ruesink2, Barzanjeh2, Seif2018}.
With blue-detuned driving, the generation of bipartite entanglement between various optical, electrical, and mechanical modes have been proposed~\cite{Vitali, Genes, Paternostro,Hofer,ChenRX,Barzanjeh, TianPRL2013,WangPRL2013, Mancini,Pirandola,Borkje,WangM,LiJ}. It was demonstrated in a recent experiment that stationary entanglement and two-mode squeezing between microwave photons can be achieved via a mechanical resonator~\cite{Barzanjeh2019}.
People have also studied the generation of tripartite entanglement via mechanical systems~\cite{Xuereb,WangPRA2015, XiangY,YangXH}. In particular, it was shown that genuine tripartite entanglement can be generated in a cavity magnomechanical system~\cite{LiPRL2018}. In experiments, entanglement between a mechanical oscillator and a microwave mode~\cite{Palomaki}, entanglement between two mechanical oscillators~\cite{Riedinger,Korppi}, and parametric amplification in mechanical and electrical modes~\cite{Massel1} have been demonstrated.

Here we present a scheme that generates switchable bipartite and genuine tripartite entanglement between microwave and optical photons via an optoelectromechanical interface. In our system, microwave and optical cavities are coupled to a mechanical resonator via radiation pressure force with the cavities driven by red- or blue-detuned fields. One of the cavities, either on the microwave or the optical side, is used as an ancilla mode to facilitate the controllability of the generated entanglement.
We show that by tuning an effective gauge phase between the linearized opto- and electro-mechanical couplings to the ``sweet spots'', bipartite entanglement can be selectively generated between output photons of designated microwave and optical cavities with the output state of the third cavity separated from the entangled state.
The entanglement of formation (EOF)~\cite{Bennett1, Bennett2}, i.e., the convex-roof extension of the entropy of entanglement, is employed to quantitatively characterize the generated continuous variable entanglement. Compared to the logarithmic negativity, an entanglement monotone used in a number of previous works~\cite{Barzanjeh,TianPRL2013,WangPRL2013}, the EOF (even though harder to compute) constitutes a proper entanglement measure satisfying properties such as convexity and asymptotic continuity that logarithmic negativity does not.
We find that the bipartite entanglement is not only switchable between different cavity outputs, but is also robust against mechanical noise and signal loss to the mechanical mode by choosing impedance-matched coupling constants.
Furthermore, using the violation of an inequality developed in \cite{Teh} and detected in \cite{ArmstrongPKLamNatPhys2015, ShalmJenneweinNatPhys2013} as the criterion, we are able to show that genuine tripartite entanglement is created between the output photons of all three cavities when the gauge phase is tuned away from the ``sweet spots''. The tripartite entanglement can be verified experimentally by conducting homodyne measurement on the quadratures of the output modes. Our result can lead to the generation and verification of multipartite entanglement in continuous variable modes with distinctively different frequencies.

The paper is organized as follows. In Sec.~\ref{sec:system}, we describe the model of the optoeletromechanical interface in detail and derive the transmission matrix connecting the input and output fields modes.
In Sec.~\ref{sec:switch}, we study the properties of the transmission matrix elements in detail and derive the conditions for achieving switchable bipartite entanglement. The bipartite entanglement is then characterized quantitatively as functions of the gauge phase, the coupling constants, the input frequency, and the thermal occupation number of the mechanical mode in Sec.~\ref{sec:bipartite}.
In Sec.~\ref{sec:tripartite}, we study the genuine tripartite entanglement when the gauge phase is tuned away from the ``sweet spots'' and discuss how to verify this entanglement by conducting homodyne detection on the quadratures of the output modes.
Conclusions are given in Sec.~\ref{sec:conclusions}.

\section{System}\label{sec:system}
\begin{figure}[h]
\includegraphics[width=7.8cm, clip]{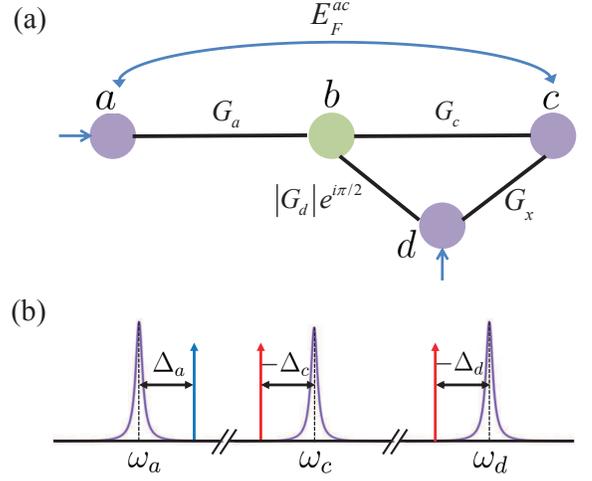}
\caption{\label{fig1} (a) The schematic of the optoelectromechanical interface. Three cavity modes $a,c,d$ are coupled to a mechanical mode $b$. The thick lines correspond to the linearized couplings $G_{a,c,d}$ and the photon hopping rate $G_{x}$. The relative phase of $G_{d}$ with respect to the other couplings is $\phi=\pi/2$.
The input fields for cavities $a$ and $d$ are indicated by arrows.
Bipartite entanglement between the outputs of cavities $a$ and $c$ is generated by mixing the inputs of $a$ and $d$ at $\phi=\pi/2$, where $E_{F}^{ac}$ represents the entanglement of formation between the outputs of cavities $a$ and $c$.
(b) Spectrum of the driving frequencies (vertical arrows) in relation to the corresponding cavity resonances $\omega_{\alpha}$. The detunings are labeled as $\Delta_\alpha$ ($\alpha=a,c,d$).}
\end{figure}

A schematic of our optoelectromechanical interface is shown in Fig.~\ref{fig1}(a). Here three cavity modes (labelled $\alpha=a,c,d$) are coupled to a common mechanical resonator $b$ via radiation pressure force. The interactions have the form $g_{\alpha} \hat{\alpha}^{\dag}\hat{\alpha} (\hat{b}+\hat{b}^{\dag})$ for cavity mode $\alpha$ with a single-photon coupling strength $g_{\alpha}$, where $\hat{\alpha}$ ($\hat{\alpha}^{\dag}$) and $\hat{b}$ ($\hat{b}^{\dag}$) are the annihilation (creation) operators of the corresponding cavity and mechanical modes.
Without loss of generality, we assume that the frequency of cavity mode $a$ is distinctively different from that of cavity modes $c$ and $d$. For example, mode $a$ could be in the microwave regime with modes $c,\,d$ in the optical regime, or vice versa.
Meanwhile, cavities $c$ and $d$ are directly coupled via a linear coupling $G_{x}(\hat{c}+\hat{c}^{\dag})(\hat{d}+\hat{d}^{\dag})$ with the coupling strength $G_{x}$.
There is no direct coupling between modes of distinctively different frequencies, such as the microwave and the optical modes, because it could result in extra circuit noise or other technical challenges.

By applying driving fields to the cavity modes, the radiation pressure interactions can be linearized. We denote $\omega_{d\alpha}$ as the driving frequency on cavity $\alpha$ and $\Delta_{\alpha}=\omega_{d\alpha}-\omega_{\alpha}$ as the detuning between the driving frequency and the cavity frequency $\omega_{\alpha}$. Assume that cavity $a$ is driven by a blue-detuned field with $\Delta_{a} \approx \omega_{m}$ and $\omega_{m}$ being the frequency of the mechanical mode, and cavities $c,\,d$ are driven by red-detuned fields with $\Delta_{c,d}\approx -\omega_{m}$, as illustrated in Fig.~\ref{fig1}(b). We can also assume that cavity $a$ is driven by a red-detuned field and cavities $c,\,d$ are driven by blue-detuned fields, and similar results can be obtained.
After a standard linearization procedure~\cite{Vitali, Genes}, the total Hamiltonian of this system in the rotating frame of the driving fields becomes $\hat{H}_{t}=\hat{H}_{0}+\hat{H}_{int}$ with
\begin{equation}
\hat{H}_{0}=-\sum_{\alpha}\Delta_{\alpha}\hat{\alpha}^{\dagger}\hat{\alpha}+\omega_{m}\hat{b}^{\dagger}\hat{b},\label{eq:H0-1}
\end{equation}
being the unperturbed Hamiltonian and $H_{int}$ being the linearized coupling Hamiltonian. Let $G_{\alpha}$ be the linearized coupling strength between cavity $\alpha$ and the mechanical mode that depends on the corresponding driving field. Here we assume $\vert G_{\alpha}\vert \ll \vert \Delta_{\alpha}\vert,\,\omega_{m}$. Under this condition and with $\vert \Delta_{\alpha}\vert\approx \omega_{m}$, we apply the rotating-wave approximation to the coupling Hamiltonian to omit fast-oscillating terms such as $\hat{a}^{\dag}\hat{b}$ and $\hat{c}\hat{b}$ in $\hat{H}_{int}$, and we have
\begin{eqnarray}
\hat{H}_{int}& =& \left(G_{a}\hat{a}\hat{b}+G_{a}^{\star}\hat{b}^{\dagger}\hat{a}^{\dag}\right)+\left(G_{c}\hat{c}^{\dagger}\hat{b}
+G_{c}^{\star}\hat{b}^{\dag}\hat{c}\right)\label{eq:Hint}\\
 &  & +\left(G_{d}\hat{d}^{\dagger}\hat{b}+G_{d}^{\star}\hat{b}^{\dag}\hat{d}\right)+\left(G_{x}\hat{c}^{\dagger}\hat{d}
 +G_{x}^{\star}\hat{d}^{\dag}\hat{c}\right),\nonumber
\end{eqnarray}
where $G_{x}$ is the photon hopping rate between cavity modes $c$ and $d$ and $G_{\alpha}^{\star}$ is the complex conjugate of the coupling constant $G_{\alpha}$. By varying the driving fields applied to the cavity modes, both the magnitude and phase of the coupling strengths $G_{a,c,d}$ can be tuned.

Besides the driving fields, the cavity and mechanical modes are also coupled to input modes $\hat{\alpha}_{\rm in}(t)$ that induce damping. The correlation functions for the cavity inputs have the form $\langle\hat{\alpha}_{\rm in}(t)\hat{\alpha}_{\rm in}^{\dag}(t^{\prime})\rangle=\delta(t-t^{\prime})$ ($\alpha=a, \,c, \,d$) at times $t$ and $t^{\prime}$, which correspond to vacuum input states. For the mechanical input field, the correlation functions are $\langle\hat{b}_{\rm in}(t)\hat{b}_{\rm in}^{\dag}(t^{\prime})\rangle=(n_{\rm th}+1)\delta(t-t^{\prime})$ and $\langle\hat{b}_{\rm in}^{\dag}(t)\hat{b}_{\rm in}(t^{\prime})\rangle=n_{\rm th}\delta(t-t^{\prime})$, which refer to thermal phonon state at temperature $T$ and thermal phonon number $n_{\rm th}=1/(e^{\hbar\omega_{m}/k_{B}T}-1)$. We only consider external damping of the cavities with damping rates $\kappa_{\alpha}$ and neglect internal dissipation for simplicity of discussion. We also assume that the mechanical damping rate $\gamma_{m}$ is much weaker than the cavity damping rates with $\gamma_{m}\ll\kappa_{\alpha}$.

The dynamics of the above interface can be described by Langevin equations.
We define a vector $\hat{v}=[\hat{a}^{\dagger},\hat{b},\hat{c},\hat{d}]^{\text{T}}$ for the system operators and $\hat{v}_{\rm in}=[\hat{a}_{\rm in}^{\dag},\hat{b}_{\rm in},\hat{c}_{\rm in},\hat{d}_{\rm in}]^{\text{T}}$ for the input operators~\cite{TianPRL2012}. Assume that the cavity detunings $\Delta_{a}=-\Delta_{c,d}=\omega_{m}$. In the rotating frame of the Hamiltonian $\hat{H}_{0}$, the Langevin equation for the vector $\hat{v}$ can be derived as
\begin{equation}
d\hat{v}/dt=M\hat{v}+\sqrt{K}\hat{v}_{\rm in},\label{eq:dvdt}
\end{equation}
with the dynamic matrix
\begin{equation}
M=\left(\begin{array}{cccc}
-\frac{\kappa_{a}}{2} & iG_{a} & 0 & 0\\
-iG_{a}^{\star} & -\frac{\gamma_{m}}{2} & -iG_{c}^{\star} & -iG_{d}^{\star}\\
0 & -iG_{c} & -\frac{\kappa_{c}}{2} & -iG_{x}\\
0 & -iG_{d} & -iG_{x}^{\star} & -\frac{\kappa_{d}}{2}
\end{array}\right)\label{eq:M}
\end{equation}
and the diagonal matrix $\sqrt{K}=\text{Diag}[\sqrt{\kappa_{a}},\sqrt{\gamma_{m}},\sqrt{\kappa_{c}},\sqrt{\kappa_{d}}]$.
This system is stable when all the real parts of the eigenvalues of the matrix $M$ are negative. Using the Routh-Hurwitz criterion~\cite{RHcriterion}, we can obtain the stability condition for our system. Denote $\Gamma_{\alpha} = 4 G_{\alpha}^{2}/\kappa_{\alpha}$. When a cavity $\alpha$ is driven by a red-detuned (blue-detuned) field with the detuning $\Delta_{\alpha}=-\omega_{m}$ ($\omega_{m}$), $\Gamma_{\alpha}$ corresponds to the cooling (heating) rate for the mechanical mode. With $\gamma_m\ll \kappa_{\alpha},G_{\alpha}$ and under the impedance matching condition $\Gamma_{c}=\Gamma_{d}$, the stability condition is $\Gamma_{c,d}> \Gamma_{a}$, i.e., the cooling rate surpasses the heating rate, as derived in Appendix~\ref{sec:stability}.

With the Fourier transformation $\hat{o}(t)=\int d\omega e^{-i\omega t}\hat{o}(\omega)/2\pi$ for an arbitrary operator $\hat{o}$, we convert the Langevin equation (\ref{eq:dvdt}) from the time domain to the frequency domain with
\begin{equation}
\hat{v}(\omega)=i(\omega I-iM)^{-1}\sqrt{K}\hat{v}_{\rm in}(\omega),\label{eq:vomega}
\end{equation}
where $I$ is the $4\times4$ identity matrix.
Here the frequency components of the system modes $\hat{v}(\omega)$ are expressed in terms of the frequency components of the input fields $\hat{v}_{\rm in}(\omega)$.
Furthermore, denote  $\hat{v}_{\rm out}=[\hat{a}_{\rm out}^{\dag},\hat{b}_{\rm out},\hat{c}_{\rm out},\hat{d}_{\rm out}]^{\text{T}}$ for the output operators. Using the input-output theorem~\cite{Gardiner} and Eq.~(\ref{eq:vomega}), we derive that $\hat{v}_{\rm out}=T(\omega)\hat{v}_{\rm in}$
with the transmission matrix
\begin{equation}
T(\omega)=I-i\sqrt{K}(\omega I-iM)^{-1}\sqrt{K}.\label{eq:wtT}
\end{equation}
The input and output operators in the frequency domain are now connected by the transmission matrix through Eq.~({\ref{eq:wtT}}). The matrix element $T_{ij}$ represents the probability amplitude of the input mode $j$ in the output mode $i$.

\section{Transmission matrix}\label{sec:switch}
In this section, we study the properties of the transmission matrix elements to understand entanglement generation between the output fields. Without loss of generality, we assume that the coupling constants $G_{a,c,x}$ are positive real numbers and $G_{d}=\vert G_{d}\vert e^{i\phi}$ is a complex number with a nontrivial phase $\phi$ (see details in Appendix~\ref{sec:phase}). The phase $\phi$ can be viewed as an effective gauge phase in the loop formed by the modes $b,c,d$ and is an important control parameter for this system.
In experiments, the linearized coupling constants $G_{\alpha}$ ($\alpha=a,c,d$) are determined by the driving field on the corresponding cavity~\cite{Vitali, Genes, Tian2010}. With current technology, the magnitude and phase of $G_{\alpha}$ can be tuned in a wide range by varying external driving fields.

To generate high-quality bipartite entanglement, it requires that (i) the input fields to be entangled are only transmitted to designated output ports (i.e., negligible loss) and (ii) the output fields in the designated ports only come from selected input ports (i.e., negligible noise). Using these requirements and considering input fields at the frequency $\omega=0$, we derive a set of operation conditions for entanglement generation: $|G_{d}|=2G_{c}G_{x}/\kappa_{c}$, $\phi=\pi/2$, and $G_{x}=\sqrt{\kappa_{c}\kappa_{d}}/2$. Details of the derivation are given in Appendix~\ref{sec:conditions}. Here $\omega=0$ in the rotating frame of $\hat{H}_{0}$ corresponds to the resonant frequency of the cavity and mechanical modes. Under these conditions, the transmission matrix can be written in terms of $\Gamma_{a,c}$ and $\gamma_{m}$ as
\begin{equation}
T=\left(\begin{array}{cccc}
-\frac{\Gamma_{c}+\Gamma_{a}+\gamma_{m}}{\Gamma_{c}-\Gamma_{a}+\gamma_{m}}
& -\frac{2i\sqrt{\Gamma_{a}\gamma_{m}}}{\Gamma_{c}-\Gamma_{a}+\gamma_{m}}
& 0
& \frac{2i\sqrt{\Gamma_{a}\Gamma_{c}}}{\Gamma_{c}-\Gamma_{a}+\gamma_{m}}\\
\frac{2i\sqrt{\Gamma_{a}\gamma_{m}}}{\Gamma_{c}-\Gamma_{a}+\gamma_{m}}
& \frac{\Gamma_{c}-\Gamma_{a}-\gamma_{m}}{\Gamma_{c}-\Gamma_{a}+\gamma_{m}}
& 0
& \frac{2\sqrt{\Gamma_{c}\gamma_{m}}}{\Gamma_{c}-\Gamma_{a}+\gamma_{m}}\\
\frac{2\sqrt{\Gamma_{a}\Gamma_{c}}}{\Gamma_{c}-\Gamma_{a}+\gamma_{m}}
& \frac{2i\sqrt{\Gamma_{c}\gamma_{m}}}{\Gamma_{c}-\Gamma_{a}+\gamma_{m}}
& 0
& \frac{-i(\Gamma_{c}+\Gamma_{a}-\gamma_{m})}{\Gamma_{c}-\Gamma_{a}+\gamma_{m}}\\
0
& 0
& i
& 0
\end{array}\right).\label{eq:Tpluspi2}
\end{equation}
Using this matrix and in the limit of $\gamma_{m}/\Gamma_{a,c} \rightarrow 0$, the outputs of cavities $a$ and $c$ can be approximated as
\begin{eqnarray}
\hat{a}_{\rm out}^{\dag} &=&
-\frac{\Gamma_{c}+\Gamma_{a}}{\Gamma_{c}-\Gamma_{a}}\hat{a}_{\rm in}^{\dag}
+ \frac{2i\sqrt{\Gamma_{a}\Gamma_{c}}}{\Gamma_{c}-\Gamma_{a}}\hat{d}_{\rm in}, \label{Eq:aout} \\
\hat{c}_{\rm out} &=&
\frac{2\sqrt{\Gamma_{a}\Gamma_{c}}}{\Gamma_{c}-\Gamma_{a}} \hat{a}_{\rm in}^{\dag}- \frac{i(\Gamma_{c}+\Gamma_{a})}{\Gamma_{c}-\Gamma_{a}}\hat{d}_{\rm in}. \label{Eq:cout}
\end{eqnarray}
These equations show that the outputs of cavities $a$ and $c$ are connected by a Bogoliubov transformation on the inputs of cavities $a$ and $d$, which clearly indicates bipartite entanglement between the output fields~\cite{TianPRL2013, WangPRL2013}. The input fields and the entanglement between the outputs of cavities $a$ and $c$ are illustrated in Fig.~\ref{fig1}(a). The matrix (\ref{eq:Tpluspi2}) also reveals the following notable features of the underlying entanglement generation process. In the limit of $\gamma_{m}\ll\Gamma_{a,c}$, the input fields of cavities $a$ and $d$ are mainly transmitted to the output fields of cavities $a$ and $c$ and become entangled. Meanwhile, the output fields of cavities $a$ and $c$ are mainly from the input fields of cavities $a$ and $d$. The input field of cavity $c$ is transmitted entirely to the output of cavity $d$. As a result, the output of cavity $d$ is not entangled with other output fields if no previous entanglement exists. And the mechanical noise is largely retained within the mechanical channel with its contribution to other output ports suppressed by a factor $\sim \sqrt{\gamma_{m}/\Gamma_{\alpha}}$. At finite temperature with thermal phonon number $n_{\rm th}$, it requires that $\gamma_{m}n_{\rm th}\ll \Gamma_{a,c}$ to make the contribution of the mechanical noise to other output ports negligible, as discussed in our previous work~\cite{JiangOE}. According to these analyses, the selected input fields have negligible loss to unwanted output ports and the designated output fields have negligible noise from unwanted input ports, especially the mechanical noise.

Meanwhile, the input and output channels of entanglement generation can be changed by choosing a different set of operation conditions. Consider the conditions $|G_{d}|=G_{c}\kappa_{d}/2G_{x}$, $\phi=-\pi/2$, and $G_{x}=\sqrt{\kappa_{c}\kappa_{d}}/2$. The transmission matrix then becomes
\begin{equation}
T=\left(\begin{array}{cccc}
-\frac{\Gamma_{c}+\Gamma_{a}+\gamma_{m}}{\Gamma_{c}-\Gamma_{a}+\gamma_{m}}
& -\frac{2i\sqrt{\Gamma_{a}\gamma_{m}}}{\Gamma_{c}-\Gamma_{a}+\gamma_{m}}
&-\frac{2\sqrt{\Gamma_{a}\Gamma_{c}}}{\Gamma_{c}-\Gamma_{a}+\gamma_{m}}
& 0 \\
\frac{2i\sqrt{\Gamma_{a}\gamma_{m}}}{\Gamma_{c}-\Gamma_{a}+\gamma_{m}}
& \frac{\Gamma_{c}-\Gamma_{a}-\gamma_{m}}{\Gamma_{c}-\Gamma_{a}+\gamma_{m}}
& \frac{2i\sqrt{\Gamma_{c}\gamma_{m}}}{\Gamma_{c}-\Gamma_{a}+\gamma_{m}}
& 0 \\
0
& 0
& 0
& i \\
-\frac{2i\sqrt{\Gamma_{a}\Gamma_{c}}}{\Gamma_{c}-\Gamma_{a}+\gamma_{m}}
& \frac{2\sqrt{\Gamma_{c}\gamma_{m}}}{\Gamma_{c}-\Gamma_{a}+\gamma_{m}}
& \frac{-i(\Gamma_{c}+\Gamma_{a}-\gamma_{m})}{\Gamma_{c}-\Gamma_{a}+\gamma_{m}}
& 0
\end{array}\right).\label{eq:Tminuspi2}
\end{equation}
In the limit of $\gamma_{m}\rightarrow0$, the output operators of cavities $a$ and $d$ can be approximated as
\begin{eqnarray}
\hat{a}_{\rm out}^{\dag} &=&
-\frac{\Gamma_{c}+\Gamma_{a}}{\Gamma_{c}-\Gamma_{a}}\hat{a}_{\rm in}^{\dag}
- \frac{2\sqrt{\Gamma_{a}\Gamma_{c}}}{\Gamma_{c}-\Gamma_{a}}\hat{c}_{\rm in}, \label{Eq:aout2} \\
\hat{d}_{\rm out} &=&
-\frac{2i\sqrt{\Gamma_{a}\Gamma_{c}}}{\Gamma_{c}-\Gamma_{a}} \hat{a}_{\rm in}^{\dag}- \frac{i(\Gamma_{c}+\Gamma_{a})}{\Gamma_{c}-\Gamma_{a}}\hat{c}_{\rm in}. \label{Eq:cout2}
\end{eqnarray}
These equations describe a Bogoliubov transformation on the inputs of cavities $a$ and $c$, which reveals the existence of bipartite entanglement between the outputs of cavities $a$ and $d$. The matrix (\ref{eq:Tminuspi2}) also shows that during the entanglement generation process, the input fields of cavities $a$ and $c$ are mainly transmitted to the output fields of cavities $a$ and $d$ to become entangled. The entanglement also exhibits robustness against both the loss of input signals to unwanted output channels and the effect of noise from unwanted input ports.

We note that both sets of operation conditions, with $\phi=\pm\pi/2$ and $G_{x}=\sqrt{\kappa_{c}\kappa_{d}}/2$, lead to the relation: $\vert G_{d}\vert=G_{c}\sqrt{\kappa_{d}/\kappa_{c}}$, which is equivalent to the impedance matching condition $\Gamma_{d}=\Gamma_{c}$ for cavity modes $c$ and $d$. In contrast, from the expression of $T_{2i}$ and $T_{i2}$ ($i=1,3,4$), we find that it requires $\gamma_{m}\ll \left(\Gamma_{c}-\Gamma_{a}\right)$ in addition to $\gamma_{m}\ll\Gamma_{a,c}$ in order for the mechanical noise not to be enhanced in the cavity outputs. Hence $\Gamma_{a}$ needs to be sufficiently smaller than $\Gamma_{c,d}$, and cavity $a$ is not impedance-matched with cavities $c$ and $d$.

\begin{figure}
\includegraphics[width=8cm]{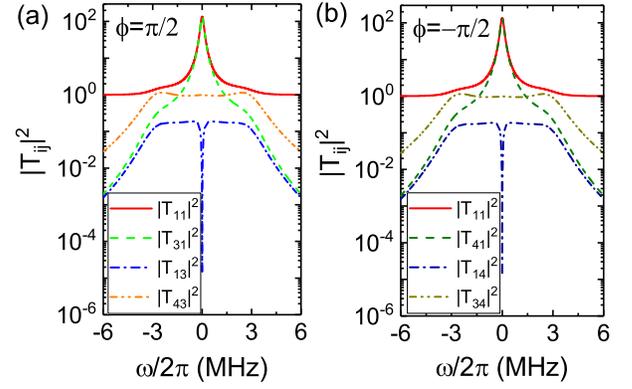}
\caption{\label{fig2} The transmission probabilities $|T_{ij}|^2$ (in logarithmic scale) as functions of the input frequency $\omega/2\pi$ at (a) $\phi=\pi/2$ and (b) $\phi=-\pi/2$. Other parameters are $\kappa_a/2\pi=2$ MHz, $\kappa_c/2\pi=3$ MHz, $\kappa_d/2\pi=3$ MHz, $\gamma_m/2\pi=0.01$ MHz, $G_a/2\pi=1.5$ MHz, $G_c/2\pi=2$ MHz, $|G_d|=G_c \sqrt{\kappa_{d}/\kappa_{c}}$, and $G_x=\sqrt{\kappa_c\kappa_d}/2$. }
\end{figure}

In Fig.~\ref{fig2}, we plot several transmission probabilities $\vert T_{ij}\vert^2$ as functions of the input frequency $\omega/2\pi$ at $\phi=\pm\pi/2$. It can be seen from Fig.~\ref{fig2}(a) that $\vert T_{11}\vert^{2}$ and $\vert T_{31}\vert^{2}$ can be much greater than unity at the phase $\phi=\pi/2$ and the input frequency $\omega=0$. Meanwhile, the probability $\vert T_{13}\vert^{2}$ approaches zero and $\vert T_{43}\vert^{2}$ approaches unity as $\omega\rightarrow 0$. This interface can hence be used as a nonreciprocal amplifier for classical input fields as studied in \cite{JiangOE}. Similar effects can be observed in Fig.~\ref{fig2}(b) as well.
The probabilities $\vert T_{11}\vert^{2}$ and $\vert T_{31}\vert^{2}$ decrease as the $\vert\omega\vert$ shifts away from the cavity resonance at $\omega=0$. The spectral halfwidth of the peaks in the transmission probabilities can be obtained from the denominators of the transmission matrix elements~\cite{TianPRL2013}. The frequency-dependent term in these denominators has the form $A(\omega)=\widetilde{\omega}_a\widetilde{\omega}_m\widetilde{\omega}_c\widetilde{\omega}_d - \widetilde{\omega}_a\left (\widetilde{\omega}_mG_x^2+\widetilde{\omega}_dG_c^2+\widetilde{\omega}_c|G_d|^2\right)  +G_a^2\left(\widetilde{\omega}_c\widetilde{\omega}_d-G_x^2\right)$ with $\widetilde{\omega}_{\alpha}=\omega_{\alpha}+i\kappa_{\alpha}/2$ ($\alpha=a,\,c,\,d$) and $\widetilde{\omega}_{m}=\omega_{m}+i\gamma_{m}/2$. Using this term, we estimate that the spectral halfwidth is on the order of magnitude of $\min[\kappa_{\alpha},\,\Gamma_{\alpha}]$. Switchable bipartite entanglement can be generated between microwave and optical photons via our interface when the frequency of the input states is within this halfwidth of the cavity resonances.

\section{Switchable bipartite entanglement}\label{sec:bipartite}
In this section, we quantitatively characterize the bipartite entanglement between designated output modes analyzed in Sec.~\ref{sec:switch}. We define a set of input and output operators:
\begin{equation}
\hat{\alpha}_x(\omega_n)=\int d\omega g_d(\omega-\omega_n)\hat{\alpha}_x(\omega)\label{eq:odiscrete}
\end{equation}
at discrete frequencies $\omega_n=n\Delta\omega$ with $n$ being an integer number, $\Delta\omega$ being a small frequency step, $\alpha=a,b,c,d$ and $x=\mathrm{in, out}$. Here $g_d(\omega)$ is a filtering function that integrates over a small frequency window with
\begin{equation}
g_d(\omega)=\left\{
\begin{array}{rr}
1/\sqrt{\Delta\omega},
& \textrm{$\omega\in(-\Delta\omega/2,\,\Delta\omega/2)$}\\
0,
& \textrm{otherwise}
\end{array}
\right.
\end{equation}
and the width of the integration window $\Delta\omega \ll\kappa_{\alpha}, g_{\alpha}$. It can be shown that the discrete input operators in (\ref{eq:odiscrete}) obey the correlation functions~\cite{TianPRL2013}
\begin{eqnarray}
&&\langle \alpha_{\mathrm{in}}(\omega_m)\alpha_{\mathrm{in}}^{\dagger}(\omega_n)\rangle=\delta_{mn}, \label{Eq:Correalpha}\\
&&\langle b_{\mathrm{in}}(\omega_m)b_{\mathrm{in}}^{\dagger}(\omega_n)\rangle=\delta_{mn}(n_{\rm th}+1), \label{Eq:CorreB}\\
&&\langle b_{\mathrm{in}}^{\dagger}(\omega_m)b_{\mathrm{in}}(\omega_n)\rangle=\delta_{mn}n_{\rm th} \label{Eq:CorreB2}
\end{eqnarray}
for the cavity modes $\alpha=a,c,d$ and the mechanical mode $b$. The correlation functions correspond to vacuum states for cavity inputs and a thermal state for the mechanical input with thermal phonon number $n_{\rm th}$. The entanglement for Gaussian states is only related to correlations of the second moments and does not depend on the first moments. For input states being coherent states of arbitrary magnitude and phase, the entanglement will be the same as that for vacuum input states.
In experiments, one can measure the discrete output modes of a cavity at selected frequencies using spectral filters.
In terms of these discrete operators, the transformation between the input and the output can be written as
$\hat{v}_{\mathrm{out}}(\omega_n) =T(\omega_n) \hat{v}_{\mathrm{in}}(\omega_n)$.

To characterize the entanglement in the cavity outputs, we denote the quadratures of the discrete output operators as
\begin{eqnarray}
\hat{X}_{\mathrm{out}}^{(\alpha)}(\omega_n)&=&\frac{1}{\sqrt{2}}[\hat{\alpha}_{\mathrm{out}}(\omega_n)+\hat{\alpha}_{\mathrm{out}}^{\dagger}(\omega_n)],\\
\hat{P}_{\mathrm{out}}^{(\alpha)}(\omega_n)&=&\frac{1}{\sqrt{2}i}[\hat{\alpha}_{\mathrm{out}}(\omega_n)-\hat{\alpha}_{\mathrm{out}}^{\dagger}(\omega_n)].
\end{eqnarray}
Define the vector $\hat{\mu}(\omega_n)=\left[\hat{X}_{\mathrm{out}}^{(a)}, \hat{P}_{\mathrm{out}}^{(a)}, \hat{X}_{\mathrm{out}}^{(c)}, \hat{P}_{\mathrm{out}}^{(c)}, \hat{X}_{\mathrm{out}}^{(d)}, \hat{P}_{\mathrm{out}}^{(d)}\right]^{\mathrm{T}}$ and the covariance matrix $V(\omega_n)$ with $V_{ij}=\langle\hat{\mu}_i (\omega_n)\hat{\mu}_j(\omega_n)+\hat{\mu}_j(\omega_n)\hat{\mu}_i(\omega_n)\rangle/2$. Based on Eqns.~(\ref{Eq:Correalpha})-(\ref{Eq:CorreB2}), we obtain
\begin{equation}
V\left(\omega_n\right)=\left(\begin{array}{ccc}
V_{aa}(\omega_n) & V_{ac}(\omega_n) & V_{ad}(\omega_n)\\
V_{ac}^{\mathrm{T}}(\omega_n) & V_{cc}(\omega_n) & V_{cd}(\omega_n) \\
V_{ad}^{\mathrm{T}}(\omega_n) & V_{cd}^{\mathrm{T}}(\omega_n) & V_{dd}(\omega_n) \\
\end{array}\right),\label{Eq:Vt}
\end{equation}
where $V_{\alpha\alpha}(\omega_n) = \mathrm{diag}[v_{\alpha\alpha},\,v_{\alpha\alpha}]$ are $2\times2$ diagonal matrices, and for $\alpha\ne\beta$,
\begin{equation}
V_{\alpha\beta}(\omega_n)=\left(\begin{array}{cc}
\mathrm{Re}[v_{\alpha\beta}]
& -\mathrm{Im}[v_{\alpha\beta}] \\
-\mathrm{Im}[v_{\alpha\beta}]
& -\mathrm{Re}[v_{\alpha\beta}]
\end{array}\right).\label{Eq:Vab}
\end{equation}
The coefficients in these matrices can be expressed as
\begin{eqnarray}
v_{aa} &=& \frac{1}{2}\left[|T_{11}|^2+(2n_{th}+1)|T_{12}|^2+|T_{13}|^2+|T_{14}|^2\right],\nonumber\\
v_{cc} &=& \frac{1}{2}\left[|T_{31}|^2+(2n_{th}+1)|T_{32}|^2+|T_{33}|^2+|T_{34}|^2\right],\nonumber\\
v_{dd} &=& \frac{1}{2}\left[|T_{41}|^2+(2n_{th}+1)|T_{42}|^2+|T_{43}|^2+|T_{44}|^2\right],\nonumber\\
v_{ac} &=& \frac{1}{2}\left[T_{11}T_{31}^*+(2n_{th}+1)T_{12}T_{32}^*+T_{13}T_{33}^*+T_{14}T_{34}^*\right],\nonumber\\
v_{ad} &=& \frac{1}{2}\left[T_{11}T_{41}^*+(2n_{th}+1)T_{12}T_{42}^*+T_{13}T_{43}^*+T_{14}T_{44}^*\right],\nonumber\\
v_{cd} &=& \frac{1}{2}\left[T_{31}T_{41}^*+(2n_{th}+1)T_{32}T_{42}^*+T_{33}T_{43}^*+T_{34}T_{44}^*\right],\nonumber
\end{eqnarray}
respectively, in terms of the transmission matrix elements and the thermal phonon number of the mechanical mode.

The covariance matrix $V(\omega_n)$ can be used to study both bipartite and tripartite entanglement in the output modes.
To calculate the bipartite entanglement between selected output modes, we use the reduced covariance matrix for these modes. Taking as example the entanglement between the outputs of cavities $a$ and $c$, the reduced covariance matrix is
\begin{equation}
V_{\mathrm{bp}}(\omega_n)=\left(\begin{array}{cc}
V_{aa}(\omega_n) & V_{ac}(\omega_n) \\
V_{ac}^{\mathrm{T}}(\omega_n) & V_{cc}(\omega_n)
\end{array}\right).\label{Eq:Vbp}
\end{equation}
By applying a phase rotation to $V_{\mathrm{bp}}(\omega_n)$, it can be transformed into the standard form $V_{\mathrm{bp}}^{\mathrm{(s)}}(\omega_n)$, where $V_{aa}$ and $V_{cc}$ remain the same as before and $V_{ac}$ becomes diagonal with $V_{ac}=\mathrm{diag}[\vert v_{ac}\vert,\,-\vert v_{ac}\vert]$.

Bipartite entanglement can be characterized by calculating the entanglement of formation (EOF)~\cite{Bennett1, Bennett2},
which quantifies the entanglement of a state in terms of the entropy of entanglement of the least entangled pure state needed to prepare it (under local operations and classical communication). Compared to logarithmic negativity, the EOF is a proper entanglement measure satisfying properties such as convexity and asymptotic continuity.
Even though, in general, EOF lacks an analytical expression, for states characterized by a covariance matrix such as
$V_{\mathrm{bp}}^{\mathrm{(s)}}(\omega_n)$, the EOF can be written as~\cite{Tserkis2017, Tserkis2019, noteEOF}
\begin{equation}
E_{F}\equiv\cosh^2r_0\log_2(\cosh^2 r_0)-\sinh^2r_0\log_2(\sinh^2 r_0),\label{eq:EOF2}
\end{equation}
where $r_{0}$  (characterizing the minimum amount of two-mode squeezing needed to create an entangled state) is given by
\begin{equation}
r_0=\frac{1}{2}\ln\sqrt{\frac{\kappa- \sqrt{\kappa ^2-\lambda_{+}\lambda_{-}}}{\lambda_{-}}}\label{eq:r0}
\end{equation}
with $\kappa=2(16\det[V_{\mathrm{bp}}^{\mathrm{(s)}}]+1)-4(v_{aa}-v_{cc})^2$ and $\lambda_{\pm}=4(v_{aa}+v_{cc} \pm 2|v_{ac}|)^{2}$. For separable states, $r_0=0$.

\begin{figure}[t]
\includegraphics[width=7.8cm]{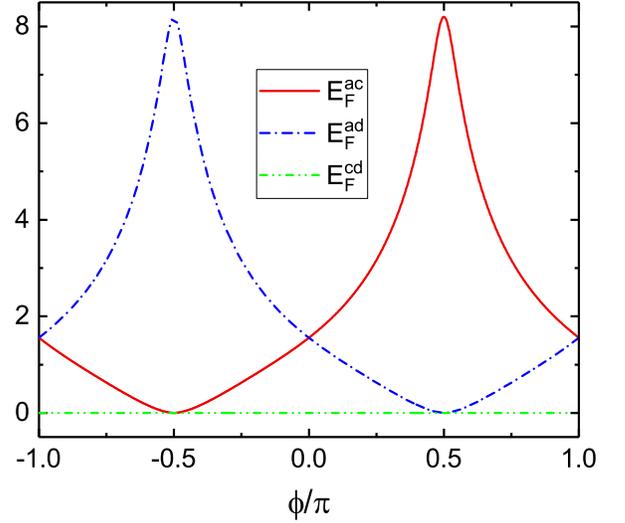}
\caption{\label{fig3} The EOFs $E_{F}^{ac}$, $E_{F}^{ad}$, and $E_{F}^{cd}$ vs the gauge phase $\phi/\pi$. Here the input frequency $\omega_{n}=0$, the thermal phonon number $n_{\mathrm{th}}=0$, and other parameters are the same as those in Fig.~\ref{fig2}.}
\end{figure}
In Fig.~\ref{fig3}, we plot the EOFs of designated cavity outputs vs the gauge phase $\phi$.
At $\phi=\pi/2$, the EOF of the outputs of cavities $a$ and $c$ reaches its maximum with $E_{F}^{ac} = 8.2$, while the EOF of the outputs of cavities $a$ and $d$ is at its minimum with $E_{F}^{ad} = 0$. The opposite can be observed at $\phi=-\pi/2$, where the outputs of cavities $a$ and $d$ are maximally entangled. In the entire range of the gauge phase, $E_{F}^{cd} = 0$, i.e., no entanglement ever exists between the outputs of cavities $c$ and $d$. This is because both the direct coupling between these two modes and their respective couplings to the mechanical mode are beam-splitter type of interaction (linear instead of bilinear).
Furthermore, at the phases $\phi=0$ and $\pm\pi$, $E_{F}^{ac} = E_{F}^{ad}$, i.e., the output of cavity $a$ is equally entangled with the outputs of cavities $c$ and $d$. This hints on the existence of nontrivial tripartite entanglement between the three output modes, which will be discussed in detail in Sec.~\ref{sec:tripartite}.
The above result shows that by choosing the gauge phase $\phi=\pm \pi/2$, we can selectively entangle the microwave photons in the output of cavity $a$ to the optical photons and switch the entanglement to either cavity $c$ or cavity $d$.

\begin{figure}[t]
\includegraphics[width=8cm]{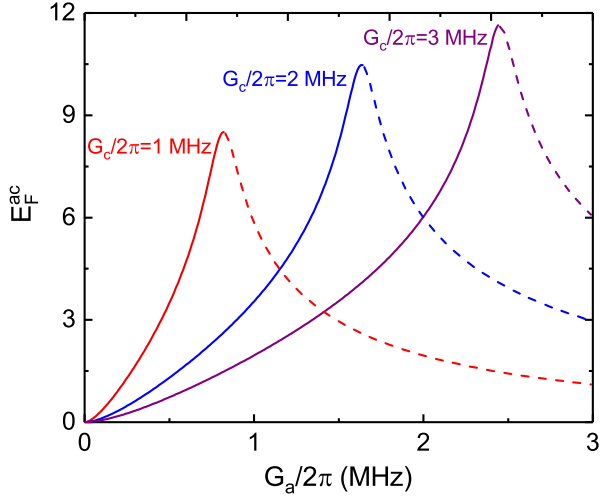}
\caption{\label{fig4} The EOF $E_{F}^{ac}$ vs the coupling $G_{a}/2\pi$ at several values of $G_c$, where the dashed lines correspond to the unstable regimes. Here $\phi=\pi/2$ and other parameters are the same as those in Fig.~\ref{fig3}.}
\end{figure}

The amount of entanglement can be tuned by varying the magnitude of the coupling constants. Based on Eqns.~(\ref{Eq:aout}) and (\ref{Eq:cout}), the coefficients in the Bogoliubov transformation on the input states diverge when $\Gamma_{a}\rightarrow\Gamma_{c}$, indicating a large amount of entanglement in the output states. To quantitatively verify this observation, we plot the EOF $E_{F}^{ac}$ as a function of $G_{a}$ at $\phi=\pi/2$ and several values of $G_{c}$ in Fig.~\ref{fig4}. It can been seen that $E_{F}^{ac}$ increases monotonically with $G_{a}$ before it reaches a maximum at $G_a\approx\sqrt{\kappa_a/\kappa_c}G_c$, i.e., $\Gamma_{a}\approx \Gamma_{c}$.
Note that as shown in Appendix~\ref{sec:stability}, the system becomes unstable when $\Gamma_{a}>\Gamma_{c}$.
On the other hand, from Eqns.~(\ref{eq:Tpluspi2}) and (\ref{eq:Tminuspi2}), we find that in the limit of $\Gamma_{a}\rightarrow\Gamma_{c}$, the mechanical noise in the outputs as well as the loss of the input states to the mechanical mode will be amplified by a factor of $2\sqrt{\Gamma_{\alpha}/\gamma_{m}}\gg1$.
Hence there is a tradeoff between generating large amount of entanglement and being robust against mechanic noise and information loss when tuning the coupling strength $G_{a}$.

\begin{figure}[t]
\includegraphics[width=8cm]{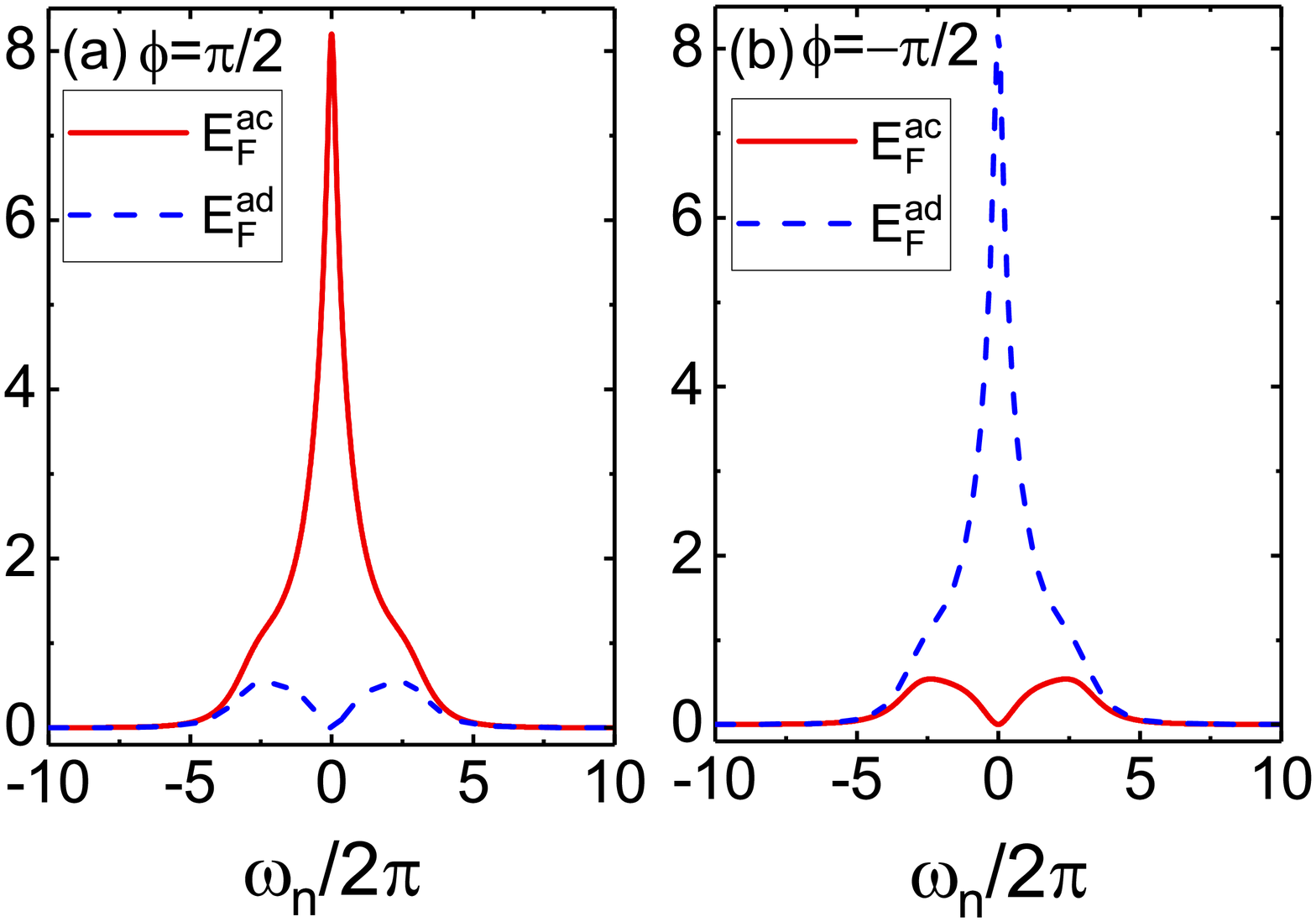}
\caption{\label{fig5} The EOFs $E_{F}^{ac}$ and $E_{F}^{ad}$ vs $\omega_{n}/2\pi$ at (a) $\phi=\pi/2$ and (b) $\phi=-\pi/2$. Other parameters are the same as those in Fig.~\ref{fig3}.}
\end{figure}

Next, we evaluate the EOF between the cavity outputs at nonzero input frequency. In Fig.~\ref{fig5}(a), we plot $E_{F}^{ac}$ and $E_{F}^{ad}$ vs the input frequency $\omega_{n}$ at $\phi=\pi/2$. Here $E_{F}^{ac}$ exhibits a maximum and $E_{F}^{ad}=0$ when the input fields are on resonance ($\omega_n=0$) with the frequencies of their corresponding cavities. The entanglement between the outputs of cavities $a$ and $c$ decreases as the input fields go off resonance. The EOF $E_{F}^{cd}\equiv0$ in the entire range of the input spectrum, agreeing with our result in Fig.~\ref{fig3}. The halfwidth of the EOF curve also agrees with our estimation in Sec.~\ref{sec:switch}.
Similarly,  by tuning the phase $\phi=-\pi/2$, the output field of cavity $a$ is entangled with the output of cavity $d$, as shown in Fig.~\ref{fig5}(b).

\begin{figure}[t]
\includegraphics[width=8cm]{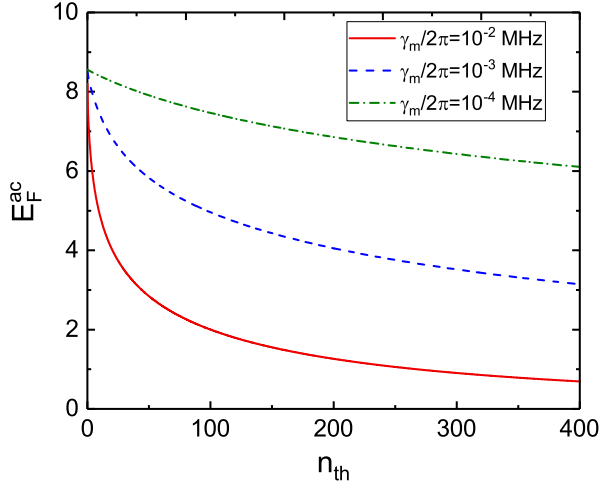}
\caption{\label{fig6} The EOF $E_{F}^{ac}$ vs the thermal phonon number $n_{\mathrm{th}}$ at several values of the mechanical damping rate $\gamma_m$. Here $\phi=\pi/2$ and other parameters are the same as those in Fig.~\ref{fig3}.}
\end{figure}

In the above discussions, we have assumed that the thermal phonon number $n_{\mathrm{th}}=0$. For a mechanical resonator with a resonant frequency $\omega_{m}/2\pi=100$ MHz, $n_{\mathrm{th}}$ will be finite even at a temperature of $20$ mK. Here we study the effect of thermal fluctuations on the bipartite entanglement generated via the optoelectromechanical interface.
In Fig.~\ref{fig6}, the EOF $E_{F}^{ac}$ is plotted vs the thermal phonon number $n_{\mathrm{th}}$ at $\phi=\pi/2$ and $\gamma_m/2\pi=10^{-2},\,10^{-3}$, and $10^{-4}$ MHz, which correspond to quality factor $Q=10^{4},\,10^{5}$, and $10^{6}$, respectively. Mechanical resonators with such quality factors have been studied in experiments. It can be seen that $E_{F}^{ac}$ decreases with the increase of $n_{\mathrm{th}}$, but the decreasing rate slows down as $n_{\mathrm{th}}$ increases. At $n_{\mathrm{th}}=400$ and a damping rate of $\gamma_m/2\pi=10^{-2}$ MHz, $E_{F}^{ac}=0.693$, which indicates the existence of entanglement between the output fields. At $n_{\mathrm{th}}=400$ and $\gamma_{m}=10^{-4}$ MHz, $E_{F}^{ac}=6.104$, which shows that a much larger entanglement can be generated as $\gamma_m$ becomes weaker~\cite{noteEOF}.
This result demonstrates the robustness of the generated continuous variable entanglement against the mechanical noise. The robustness  is rooted in the choice of the operation conditions discussed in Appendix~\ref{sec:conditions} with the mechanical noise effectively retained to tshe mechanical channel.

\section{Genuine tripartite entanglement}\label{sec:tripartite}
Multipartite entanglement is crucial for quantum communication between multiparties in a quantum network~\cite{Horodecki}. The generation of multipartite entanglement in hybrid quantum systems with distinctively different frequencies is often hindered by noise propagation, signal loss and the requirement on the couplings. In this section, we will show that it is possible to generate genuine tripartite entanglement between microwave and optical photons in the cavity outputs via our optoelectromechanical interface.

A three-mode system is genuinely tripartite entangled when the density matrix of the system cannot be decomposed into a mixture of bi-separable states. Criteria have been developed in previous works to verify the existence of genuine tripartite entanglement in continuous variable systems~\cite{Teh}.
A sufficient (but not necessary) criterion for genuine tripartite entanglement is the violation of the inequality
\begin{align}
\Delta \hat{u}\Delta \hat{v}\geq&\min\{|g_3 h_3|+|h_1 g_1+h_2 g_2|,\nonumber\\&|g_2 h_2|+|h_1 g_1+h_3 g_3|,\nonumber\\&|g_1 h_1|+|g_2 h_2+h_3 g_3|\},\label{eq:inequality}
\end{align}
where $\Delta \hat{u}$ ($\Delta \hat{v}$) is the variance of the operator $\hat{u}$ ($\hat{v}$) with
\begin{eqnarray}
\hat{u}=h_1 \hat{X}_\mathrm{out}^{(a)}+h_2 \hat{X}_\mathrm{out}^{(c)}+h_3 \hat{X}_\mathrm{out}^{(d)} \label{eq:u}\\
\hat{v}=g_1 \hat{P}_\mathrm{out}^{(a)}+g_2 \hat{P}_\mathrm{out}^{(c)}+g_3 \hat{P}_\mathrm{out}^{(d)} \label{eq:v}
\end{eqnarray}
defined in terms of the quadratures of the cavity outputs $\hat{X}_\mathrm{out}^{(\alpha)}$ and $\hat{P}_\mathrm{out}^{(\alpha)}$ and
$g_i$ and $h_i$ ($i=1,2,3$) being real numbers. When this inequality can be violated for arbitrary choices of $g_i$ and $h_i$, the system exhibits genuine tripartite entanglement.

\begin{figure}[t]
\includegraphics[width=8cm]{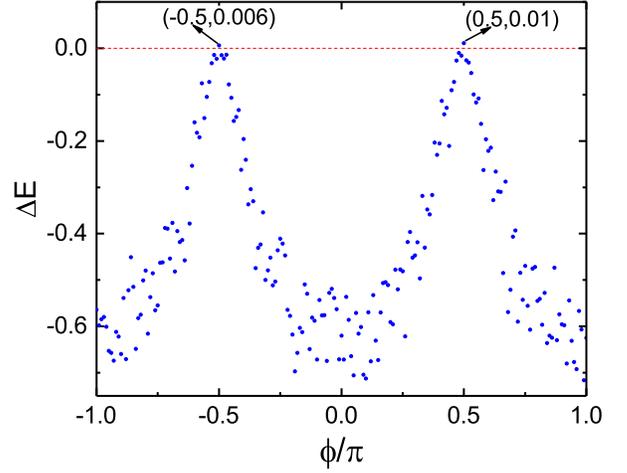}
\caption{\label{fig7} The difference $\Delta E$ vs the relative phase $\phi$. Other parameters are the same as those in Fig.~\ref{fig3}.}
\end{figure}
To test the above inequality, we define the difference $\Delta E=\Delta \hat{u}\Delta \hat{v}-\min\{|g_3 h_3|+|h_1 g_1+h_2 g_2|,|g_2 h_2|+|h_1 g_1+h_3 g_3|,|g_1 h_1|+|g_2 h_2+h_3 g_3|\}$ and choose $5\times10^4$ random numbers for each of $g_i$ and $h_i$ with $\{g_i,\,h_i\}\in[-1,1]$. Even though we cannot conduct the test on arbitrary $g_i$ and $h_i$, we think this is a sufficiently large pool of choices for our purpose. We then calculate $\Delta E$ for these random numbers at $\omega_{n}=0$, $n_{\mathrm{th}}=0$, and the gauge phase $\phi$. In Fig.~\ref{fig7}, $\Delta E$ is plotted as a function of $\phi$.
It can be seen that the inequality (\ref{eq:inequality}) is always violated at $\phi\neq\pm\pi/2$ with all $\Delta E<0$, which
provides evidence for genuine tripartite entanglement in the cavity outputs.
On the contrary, at the ``sweet spots'' of $\phi=\pm\pi/2$, there exist sets of $g_i$ and $h_i$ with $\Delta E>0$. We also observe that at $\phi=\pi/2$, the matrix elements $v_{ad}=v_{cd}=0$, i.e., the output state of cavity $d$ is separable from the output state of cavities $a$ and $c$, and hence there is no genuine tripartite entanglement. Similar result can be found at $\phi=-\pi/2$.
This agrees with our result in Fig.~\ref{fig3}, where bipartite entanglement exists only between the outputs of cavities $a$ and $c$ ($d$) with the output of cavity $d$ ($c$) separable from the states of the other modes at $\phi=\pi/2$ ($-\pi/2$).
We also note that $\vert \Delta E\vert$ (or the violation) is at its maximum when the gauge phase approaches $\phi=0$.
Although the degree of violation of the inequality (\ref{eq:inequality}) does not constitute an entanglement measure of genuine tripartite entanglement, it indicates that $\phi=0$ is probably where it is easiest to observe such entanglement.

\begin{figure}[t]
\includegraphics[width=8cm]{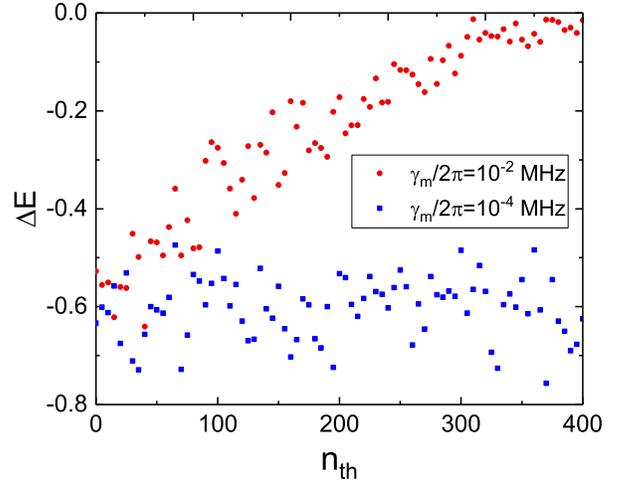}
\caption{\label{fig8} The difference $\Delta E$ vs the thermal phonon number $n_{\mathrm{th}}$ at the gauge phase $\phi=0$. Other parameters are the same as those in Fig.~\ref{fig3}.}
\end{figure}
In Fig.~\ref{fig8}, we plot $\Delta E$ as a function of the thermal phonon number $n_{\mathrm{th}}$ at $\phi=0$ and $\omega_n=0$. It can be seen that $\Delta E$ remains finite even at a thermal phonon number of $n_{\mathrm{th}}=400$ and a mechanical damping rate of $\gamma_m/2\pi=10^{-2}$ MHz. The difference $\Delta E$ becomes even larger with $\Delta E\sim -0.6$ at $n_{\mathrm{th}}=400$ and a much weaker damping rate of $\gamma_m/2\pi=10^{-4}$ MHz. This result shows that the genuine tripartite entanglement generated in our system is robust against thermal noise, similar to the behavior of the bipartite entanglement shown in Fig.~\ref{fig6}.

An interesting effect in our system is that the outputs of cavities $c$ and $d$ are never entangled bipartitely with $E_{F}^{cd}\equiv 0$ regardless of the gauge phase or the input frequency, as has been shown in Sec.~\ref{sec:bipartite}. But this does not prevent our system from being in a genuine tripartite entangled state due to the entanglement between the outputs of cavities $a$ and $c$ and between the outputs of cavities $a$ and $d$.  Similar phenomenon has been explored in a cavity magnomechanical system~\cite{LiPRL2018}, where it was shown that genuine tripartite entanglement can exist even in the case that one bipartite entanglement is absent.

The genuine tripartite entanglement in our system can be verified by measuring the variances and covariances of the quadratures of the cavity outputs. With spectral filtering of the output modes and homodyne detection on the filtered states, the matrix elements of the covariance matrix $V(\omega_{n})$, and subsequently the difference $\Delta E$, can be obtained. Such homodyne detection can be readily performed in the optical domain, and it has also been demonstrated for microwave photons in recent experiments on superconducting systems~\cite{Wallraff2013}.

\section{Conclusions}\label{sec:conclusions}
To summarize, we presented a scheme to generate switchable bipartite and genuine tripartite entanglement between microwave and optical photons via an optoelectromechanical interface. The bipartite entanglement can be generated in designated output channels by manipulating an effective gauge phase between the linearized opto- and electro-mechanical couplings. We characterized the entanglement quantitatively with the EOF and showed that the entanglement is robust against mechanical noise and signal loss to the mechanical mode. We also revealed the generation of genuine tripartite entanglement through the violation of an inequality. The tripartite entanglement can be verified  experimentally by performing homodyne detections on the quadratures of the output modes.
Our result can lead to future studies of entanglement and quantum communication via mechanical interfaces in multipartite hybrid systems.

\section*{ACKNOWLEDGMENTS}
The authors would like to thank Prof. Timothy C. Ralph for valuable discussions. C.J. was supported by the Natural Science Foundation of China (NSFC) under Grant No. 11874170, the Postdoctoral Science Foundation of China under Grant No. 2017M620593, and Qing Lan Project of Universities in Jiangsu Province. Y.L. was supported by the National Natural Science Foundation of China (under Grants No.~11774024, No.~11534002). S.T. and S.O. are supported by the Australian Research Council (ARC) under the Centre of Excellence for Quantum Computation and Communication Technology (Grant No. CE170100012). K.C. and L.T. are supported by the National Science Foundation (USA) under Award No. PHY-1720501 and the UC Multicampus-National Lab Collaborative Research and Training under Award No. LFR-17-477237.

\appendix
\section{Couplings and gauge phase}\label{sec:phase}
In Sec.~\ref{sec:switch}, we assume that the coupling constants $G_{a,c,x}$ are positive real numbers and $G_{d}=\vert G_{d}\vert e^{i\phi}$ is a complex number with nontrivial phase $\phi$, which corresponds to a gauge phase. Here we explain the origin of the gauge phase and show that this assumption does not cause any loss of generality. Let all the couplings be complex numbers with $G_{\alpha}=\vert G_{\alpha}\vert^{i\phi_\alpha}$ ($\alpha=a,c,d,x$) initially. We redefine the operators as:
\begin{eqnarray}
\hat{a} & \rightarrow & \hat{a}, \label{eq:asub} \\
\hat{b} & \rightarrow & \hat{b} e^{-i\phi_a}, \label{eq:bsub} \\
\hat{c} & \rightarrow & \hat{c} e^{i\left(\phi_{c}-\phi_a\right)}, \label{eq:csub} \\
\hat{d} & \rightarrow & \hat{d} e^{i\left(-\phi_{x}+\phi_{c}-\phi_a\right)}. \label{eq:dsub}
\end{eqnarray}
The creation operators are redefined accordingly. With these definitions, the Hamiltonian (\ref{eq:Hint}) becomes
\begin{eqnarray}
\hat{H}_{int}& =& \left(\vert G_{a}\vert \hat{a}\hat{b}+\vert G_{a}\vert\hat{b}^{\dagger}\hat{a}^{\dag}\right)+\left(\vert G_{c}\vert \hat{c}^{\dagger}\hat{b} +\vert G_{c}\vert\hat{b}^{\dag}\hat{c}\right)\label{eq:Hintredefined}\\
 &  & +\left(\vert G_{d}\vert e^{i\phi}\hat{d}^{\dagger}\hat{b}+\vert G_{d}\vert e^{-i\phi}\hat{b}^{\dag}\hat{d}\right)+\left(\vert G_{x}\vert \hat{c}^{\dagger}\hat{d} +\vert G_{x}\vert \hat{d}^{\dag}\hat{c}\right)\nonumber
\end{eqnarray}
with $\phi=\phi_{d}+\phi_{x}-\phi_{c}$. Hence by redefining the operators, the coupling constants $G_{a,c,x}\rightarrow \vert G_{a,c,x}\vert$, becoming positive real numbers, and $G_{d}\rightarrow \vert G_{d}\vert e^{i\phi}$. Basically, the phase factors of the complex couplings are now absorbed into the definition of the operators, which leaves only a nontrivial phase $\phi$ in the coupling $G_{d}$. This phase is the accumulated phase of the couplings in the loop composed of modes $b,c,d$ and can be treated as a gauge phase of the loop.

In experiments, the linearized coupling constants $G_{\alpha}$ ($\alpha=a,c,d$) are determined by the driving field on the corresponding cavity mode. As shown in previous works~\cite{Vitali, Genes, Tian2010}, $G_{\alpha}$ is linearly proportional to the driving field. By adjusting the strength and phase of the driving field, the magnitude and phase of $G_{\alpha}$ can be tuned. With current technology, $G_{\alpha}$ can be tuned in a wide range with designated phase. Therefore, the gauge phase can be well controlled via external driving fields.

\section{Operation conditions}\label{sec:conditions}
Here we utilize two requirements on entanglement generation to derive the operation conditions. Following Appendix~\ref{sec:phase}, $G_{a,c,x}$ are positive numbers and $G_{d}=\vert G_{d}\vert e^{i\phi}$. And we consider the case of $\omega=0$.

\emph{(i) The input fields to be entangled are only transmitted to designated output ports (i.e., negligible loss).} This requirement ensures that the signal fields are not transmitted to other output ports besides the designated ones. Consider the transmission of the input field $\hat{a}_{in}$, which is characterized by the matrix elements $T_{i1}$ with $i=1,2,3,4$. Using Eq.~(\ref{eq:wtT}), we have
\begin{equation}
\frac{T_{41}}{T_{31}} = \sqrt{\frac{\kappa_{d}}{\kappa_{c}}}\frac{G_{d}\kappa_{c}-2iG_{c}G_{x}}{G_{c}\kappa_{d}-2iG_{d}G_{x}}. \label{eq:T41T31}
\end{equation}
By choosing
\begin{equation}
|G_{d}|=2G_{c}G_{x}/\kappa_{c}\,\, {\rm and}\,\, \phi=\pi/2, \label{eq:condition1}
\end{equation}
we find
\begin{eqnarray}
\frac{T_{11}}{T_{31}}&=&-\frac{4G_{c}^{2}\kappa_{a}+4G_{a}^{2}\kappa_{c}+\gamma_{m}\kappa_{a}\kappa_{c}}{8G_{a}G_{c}\sqrt{\kappa_{a}\kappa_{c}}},\label{eq:T11T31} \\
\frac{T_{21}}{T_{31}}&=&\frac{i\sqrt{\gamma_{m}\kappa_{c}}}{2G_{c}},\label{eq:T21T31} \\
\frac{T_{41}}{T_{31}}&=&0.\label{eq:T41T31simple}
\end{eqnarray}
In the limit of $\gamma_m\ll \Gamma_{a,c}$, Eq.~(\ref{eq:T11T31}) shows that $\vert T_{11}/T_{31}\vert >1$ and approaches unity only when $\gamma_{m}=0$ and $\Gamma_{a}=\Gamma_{c}$. Eq.~(\ref{eq:T21T31}) shows that $\vert T_{21}/T_{31}\vert \ll 1$. Therefore the input field of cavity $a$ is mainly transmitted to the outputs of cavities $a$ and $c$ with no or negligible contribution to the outputs of cavity $d$ and mechanical mode $b$. This result also identifies the outputs of cavities $a$ and $c$ as the designated output ports.

\emph{(ii) The output fields in the designated ports only come from selected input ports (i.e., negligible noise).} This requirement ensures that the output fields do not have contributions from other input ports besides the selected ones. Consider the output field of cavity $c$ with $\hat{c}_{\rm out}=\sum_{i} T_{3i}\hat{v}_{\rm in,i}$ ($i=1,2,3,4$). In addition to the condition (\ref{eq:condition1}), by choosing \begin{equation}
G_{x}=\sqrt{\kappa_{c}\kappa_{d}}/2, \label{eq:condition2}
\end{equation}
we have
\begin{eqnarray}
\frac{T_{32}}{T_{31}}&=&\frac{i\sqrt{\gamma_{m}\kappa_{a}}}{2G_{a}}, \label{eq:T32T31}\\
\frac{T_{33}}{T_{31}}&=&0, \label{eq:T33T31} \\
\frac{T_{34}}{T_{31}}&=&-\frac{i\left(4G_{c}^{2}\kappa_{a}+4G_{a}^{2}\kappa_{c}-\gamma_{m}\kappa_{a}\kappa_{c}\right)}{8G_{a}G_{c}\sqrt{\kappa_{a}\kappa_{c}}}.\label{eq:T34T31}
\end{eqnarray}
The output of cavity $c$ thus includes components from the inputs of cavities $a$ and $d$, but not from the input of cavity $c$. With $\gamma_m\ll \Gamma_{a,c}$, $\vert T_{32}/T_{31}\vert \ll 1$, i.e., the mechanical noise has negligible contribution to the output of cavity $c$.

In the above analyses, we apply the negligible loss requirement on the input of cavity $a$ and the negligible noise requirement on the output of cavity $c$ to derive a set of operation conditions (\ref{eq:condition1}) and (\ref{eq:condition2}) for entanglement generation. Under these conditions, $T_{14}=iT_{31}$, $T_{34}\approx iT_{11}$, $\vert T_{24}/T_{31}\vert=\sqrt{\gamma_{m}/\Gamma_{a}}\ll1$, and $T_{44}=0$. Hence the input field of cavity $d$ is mainly transmitted to the outputs of cavities $a$ and $c$ with the negligible loss requirement also satisfied for the input field of cavity $d$. We also find that $T_{12}=-T_{21}$, $T_{13}=0$, and $T_{14}=iT_{31}$. It can be seen that the output field of cavity $a$ mainly includes contributions from the inputs of cavities $a$ and $d$ with the negligible noise requirement satisfied for the output field of cavity $a$ as well. Furthermore, $T_{43}=i$ and $T_{4i}=0$ ($i=1,2,4$), i.e., the output of cavity $d$ is only from the input of cavity $c$. While $T_{22}\approx 1$ and $\vert T_{i2}/T_{31}\vert\ll 1$ ($i=1,3,4$), which shows that the mechanical noise is mainly retained in the mechanical channel. Hence under the conditions (\ref{eq:condition1}) and (\ref{eq:condition2}), the input fields of cavities $a$ and $d$ are transmitted to the output fields of cavities $a$ and $c$ to become entangled. The entanglement generation has negligible loss and negligible noise from the mechanical mode.

Under the operation conditions, the expression of the transmission matrix elements can be simplified using $\Gamma_{a,c,d}$ and the $\gamma_{m}$. For example, (\ref{eq:T11T31}) can be written as
\begin{equation}
\frac{T_{11}}{T_{31}}=-\frac{\Gamma_{a}+\Gamma_{c}+\gamma_{m}}{2\sqrt{\Gamma_{a}\Gamma_{c}}}.\label{eq:T11T31v2}
\end{equation}

Meanwhile, the input and output channels of entanglement generation can be changed by choosing a different set of operation conditions. When $|G_{d}|=G_{c}\kappa_{d}/2G_{x}$, $\phi=-\pi/2$, and $G_{x}=\sqrt{\kappa_{c}\kappa_{d}}/2$, it can be shown that the input fields of cavities $a$ and $c$ are transmitted to the output fields of cavities $a$ and $d$ to become entangled.

\section{Stability condition}\label{sec:stability}
With blue-detuning driving on (at least) one of the cavities, the optoelectromechanical interface can become unstable. Here we use the Routh-Hurwitz criterion to study the stability of this system~\cite{RHcriterion}. When the real parts of all four eigenvalues of the dynamic matrix $M$ in Eq.~(\ref{eq:M}) are negative, the system is in the stable regime.
It can be shown that the eigenvalues of the matrix $M$ satisfy the equation:
$\lambda^4+s_3\lambda^3+s_2\lambda^2+s_1\lambda+s_0=0$
with the coefficients
\begin{eqnarray}
s_3=&&\left(\kappa_a+\kappa_c+\kappa_d+\gamma_m\right)/2,\\
s_2=&&\left[\kappa_a\kappa_c+\kappa_a\kappa_d+\kappa_c\kappa_d+
\gamma_m \left(\kappa_a+\kappa_c+\kappa_d\right)\right]/4\nonumber\\
 &&+G_x^2+G_c^2+|G_d|^2-G_a^2,\\
s_1=&&\left[\gamma_m\left(\kappa_a\kappa_c+\kappa_a\kappa_d+\kappa_c\kappa_d\right)
+\kappa_a\kappa_c\kappa_d\right]/8\nonumber\\
&&+\left(\kappa_a+\gamma_m\right)G_x^2/2+\left(\kappa_a+\kappa_d\right)G_c^2/2\nonumber\\
&&+\left(\kappa_a+\kappa_c\right)|G_d|^2/2 -\left(\kappa_c+\kappa_d\right)G_a^2/2,\\
s_0=&&\kappa_a\gamma_m\kappa_c\kappa_d/16+\kappa_a\gamma_m G_x^2/4
+\kappa_a\kappa_d G_c^2/4 \nonumber\\
&&+\kappa_a\kappa_c|G_d|^2/4-\kappa_c\kappa_d G_a^2/4-G_a^2 G_x^2.
\end{eqnarray}
We find that for the system to be stable, these coefficients should satisfy the following relations: (1) all $s_i>0$, (2) $s_3s_2-s_1>0$, and (3) $s_3s_2s_1-s_1^2-s_0s_3^2>0$. Under the conditions of $\gamma_m\ll \kappa_{\alpha}, G_{\alpha}$ and $\Gamma_{c}=\Gamma_{d}$, these relations can be approximated as $\Gamma_{c,d}>\Gamma_{a}$.


\begin{thebibliography}{99}

\bibitem{Horodecki}R. Horodecki, P. Horodecki, M. Horodecki, and K. Horodecki, Rev. Mod. Phys. \textbf{81}, 865 (2009).

\bibitem{Kimble}H. J. Kimble, Nature \textbf{453,} 1023 (2008).

\bibitem{Berkley}A. J. Berkley, H. Xu, R. C. Ramos, M. A. Gubrud, F. W. Strauch, P. R. Johnson, J. R. Anderson, A. J. Dragt, C. J. Lobb, F. C. Wellstood, Science \textbf{300} 1548 (2003).

\bibitem{Steffen}M. Steffen, M. Ansmann, R. C. Bialczak, N. Katz, E. Lucero, R. McDermott, Matthew Neeley, E. M. Weig, A. N. Cleland, and J. M. Martinis, Science \textbf{313}, 1423 (2006).

\bibitem{Julsgaard}B. Julsgaard, A. Kozhekin, and E. S. Polzik, Nature \textbf{413}, 400 (2001).

\bibitem{ArmstrongPKLamNatPhys2015} S. Armstrong, M. Wang, R. Y. Teh, Q. Gong, Q. He, J. Janousek, H.-A. Bachor, M. D. Reid, and P. K. Lam, Nat. Phys. \textbf{11}, 167 (2015).

\bibitem{ShalmJenneweinNatPhys2013} L. K. Shalm, D. R. Hamel, Z. Yan, C. Simon, K. J. Resch and T. Jennewein, Nat. Phys. \textbf{9}, 19 (2013).

\bibitem{Tian2004} L. Tian, P. Rabl, R. Blatt, and P. Zoller, Phys. Rev. Lett. \textbf{92}, 247902 (2004).

\bibitem{Stannigel2010} K. Stannigel, P. Rabl, A. S. S{\o}rensen, P. Zoller, and M. D. Lukin, Phys. Rev. Lett. \textbf{105}, 220501 (2010).

\bibitem{Metcalfe2014}M. Metcalfe, Appl. Phys. Rev. \textbf{1}, 031105 (2014).

\bibitem{Tian2015} L. Tian, Ann. Phys. (Berlin) \text{527}, 1 (2015).

\bibitem{WangPRL2012}Y. D. Wang and A. A. Clerk, Phys. Rev. Lett. \textbf{108,} 153603 (2012).

\bibitem{TianPRL2012}L. Tian, Phys. Rev. Lett. \textbf{108,} 153604 (2012).

\bibitem{BarzanjehPRL2012}S. Barzanjeh, M. Abdi, G. J. Milburn, P. Tombesi, and D. Vitali, Phys. Rev. Lett. \textbf{109,} 130503 (2012).

\bibitem{AspelmeyerRMP}M. Aspelmeyer, T. J. Kippenberg, and F. Marquardt, Rev. Mod. Phys. \textbf{86}, 1391 (2014).

\bibitem{Weis} S. Weis, R. Rivi\`{e}re , S. Del\'{e}glise, E. Gavartin, O. Arcizet, A. Schliesser, and T. J. Kippenberg, Science \textbf{330}, 1520 (2010).

\bibitem{Naeini1} A. H. Safavi-Naeini, T. P. Mayer Alegre, J. Chan, M. Eichenfield, M. Winger, Q. Lin, J. T.Hill, D. E. Chang, and O. Painter, Nature \textbf{472}, 69 (2011).

\bibitem{Chan}J. Chan, T. P. M. Alegre, A. H. Safavi-Naeini, J. T. Hill, A. Krause, S. Gr\"{o}blacher, M. Aspelmeyer, and O. Painter, Nature \textbf{478}, 89 (2011).

\bibitem{Teufel}J. D. Teufel, T. Donner, D. Li, J. W. Harlow, M. S. Allman, K. Cicak, A. J. Sirois, J. D. Whittaker, K. W. Lehnert, and R. W. Simmonds, Nature (London) \textbf{475}, 359 (2011).

\bibitem{Andrews2014}R. W. Andrews, R. W. Peterson, T. P. Purdy, K. Cicak, R. W. Simmonds, C. A. Regal, and K. W. Lehnert, Nat. Phys. \textbf{10} 321 (2014).

\bibitem{Cleland2013} J. Bochmann, A. Vainsencher, D. D. Awschalom, and A. N. Cleland, Nat. Phys. \textbf{9}, 712 (2013).

\bibitem{Polzik2014} T. Bagci, A. Simonsen, S. Schmid, L. G. Villanueva, E. Zeuthen, J. Appel, J. M. Taylor, A. Sorensen, K. Usami, A. Schliesser, and E. S. Polzik, Nature (London) \textbf{507}, 81 (2014).

\bibitem{Hafezi}M. Hafezi and P. Rabl, Opt. Express \textbf{20}, 7672 (2012).

\bibitem{ShenZ}Z. Shen, Y.-L. Zhang, Y. Chen, C.-L. Zou, Y.-F. Xiao, X.-B. Zou, F.-W. Sun, G.-C. Guo, and C.-H. Dong, Nat. Photon. \textbf{10}, 657 (2016).

\bibitem{Ruesink}F. Ruesink, M.-A. Miri, A. Al\`{u}, and E. Verhagen, Nat. Commun. \textbf{7}, 13662 (2016).

\bibitem{Miri}M.-A. Miri, F. Ruesink, E. Verhagen, and A. Al\`{u}, Phys. Rev. Appl. \textbf{7}, 064014 (2017).

\bibitem{TianPRA2017}L. Tian and Z. Li, Phys. Rev. A \textbf{96}, 013808 (2017).

\bibitem{XuXW2}X.-W. Xu, Y. Li, A.-X. Chen, and Y.-x. Liu, Phys. Rev. A \textbf{93}, 023827 (2016).

\bibitem{Bernier}N. R. Bernier, L. D. T\'{o}th, A. Koottandavida, M. A. Ioannou, D. Malz, A. Nunnenkamp, A. K. Feofanov, and T. J. Kippenberg, Nat. Commun. \textbf{8}, 604 (2017);

\bibitem{Barzanjeh2017} S. Barzanjeh, M. Wulf, M. Peruzzo, M. Kalaee, P. B. Dieterle, O. Painter, and J. M. Fink, Nat. Commun. \textbf{8}, 953 (2017).

\bibitem{Teufel2017} G. A. Peterson, F. Lecocq, K. Cicak, R. W. Simmonds, J. Aumentado, and J. D. Teufel, Phys. Rev. X \textbf{7}, 031001 (2017).

\bibitem{FangKJ} K. Fang, J. Luo, A. Metelmann, M. H. Matheny, F. Marquardt, A. A. Clerk, and O. Painter, Nat. Phys. \textbf{13}, 465 (2017).

\bibitem{Malz} D. Malz, L. D. T\'{o}th, N. R. Bernier, A. K. Feofanov, T. J. Kippenberg, A. Nunnenkamp, Phys. Rev. Lett. \textbf{120}, 023601 (2018).

\bibitem{Sillanpaa2018}L. Mercier de L\'{e}pinay, E. Damsk\"{a}gg, C. F. Ockeloen-Korppi, M. A. Sillanp\"{a}\"{a}, Phys. Rev. Applied \textbf{11}, 034027 (2019).

\bibitem{LiY}Y. Li, Y. Y. Huang, X. Z. Zhang, and L. Tian, Opt. Express \textbf{25}, 18907 (2017).

\bibitem{ZhangXZ}X. Z. Zhang, L. Tian, and Y. Li, Phys. Rev. A \textbf{97}, 043818 (2018).

\bibitem{JiangOE}C. Jiang, B. W. Ji, Y. S. Cui, F. Zuo, J. Shi, G. B. Chen, Opt. Express, \textbf{26}, 15255 (2018).

\bibitem{ShenZ2}Z. Shen, Y.-L. Zhang, Y. Chen, F.-W. Sun, X.-B. Zou, G.-C. Guo, C.-L. Zou, C.-H. Dong, Nat. Commun. \textbf{9}, 1797 (2018).

\bibitem{Ruesink2}F. Ruesink, J. P. Mathew, M.-A. Miri, A. Al\`{u}, and E. Verhagen, Nat. Commun. \textbf{9}, 1798 (2018).

\bibitem{Barzanjeh2}S. Barzanjeh, M. Aquilina, and A. Xuereb, Phys. Rev. Lett. \textbf{120}, 060601 (2018).

\bibitem{Seif2018} A. Seif, W. DeGottardi, K. Esfarjani, and M. Hafezi, Nat. Commun. \textbf{9}, 1207 (2018).

\bibitem{Vitali}D. Vitali, S. Gigan, A. Ferreira, H. R. B\"{o}hm, P. Tombesi, A. Guerreiro, V. Vedral, A. Zeilinger, and M. Aspelmeyer, Phys. Rev. Lett. \textbf{98}, 030405 (2007).

\bibitem{Genes}C. Genes, A. Mari, P. Tombesi, and D. Vitali, Phys. Rev. A \textbf{78}, 032316 (2008).

\bibitem{Paternostro} M. Paternostro, D. Vitali, S. Gigan, M. S. Kim, C. Brukner, J. Eisert, and M. Aspelmeyer, Phys. Rev. Lett. \textbf{99}, 250401 (2007).

\bibitem{Hofer}S. G. Hofer, W. Wieczorek, M. Aspelmeyer, and K. Hammerer, Phys. Rev. A \textbf{84}, 052327 (2011).

\bibitem{ChenRX}R. X. Chen, L. T. Shen, and S. B. Zheng, Phys. Rev. A \textbf{91}, 022326 (2015).

\bibitem{Barzanjeh}S. Barzanjeh, D. Vitali, P. Tombesi, and G. J. Milburn, Phys. Rev. A \textbf{84}, 042342 (2011).

\bibitem{TianPRL2013}L. Tian, Phys. Rev. Lett. \textbf{110}, 233602 (2013).

\bibitem{WangPRL2013}Y. D. Wang and A. A. Clerk, Phys. Rev. Lett. \textbf{110}, 253601 (2013).

\bibitem{Mancini}S. Mancini, V. Giovannetti, D. Vitali, and P. Tombesi, Phys. Rev. Lett. \textbf{88}, 120401 (2002).

\bibitem{Pirandola}S. Pirandola, D. Vitali, P. Tombesi, and S. Lloyd, Phys. Rev. Lett. \textbf{97}, 150403 (2006).

\bibitem{Borkje}K. B{\o}rkje, A. Nunnenkamp, and S. M. Girvin, Phys. Rev. Lett. \textbf{107}, 123601 (2011).

\bibitem{WangM}M. Wang, X. Y. L\"{u}, Y. D. Wang, J. Q. You, and Y. Wu, Phys. Rev. A \textbf{94}, 053807 (2016).

\bibitem{LiJ}J. Li, G. Li, S. Zippilli, D. Vitali, and T. C. Zhang, Phys. Rev. A \textbf{95}, 043819 (2017).

\bibitem{Barzanjeh2019}S. Barzanjeh, E. S. Redchenko, M. Peruzzo, M. Wulf, D. P. Lewis, G. Arnold. \& J. M. Fink, Nature (London) 570, 480 (2019).

\bibitem{Xuereb}A. Xuereb, M. Barbieri, and M. Paternostro, Phys. Rev. A \textbf{86}, 013809 (2012).

\bibitem{WangPRA2015}Y. D. Wang, S. Chesi, and A. A. Clerk, Phys. Rev. A \textbf{91}, 013807 (2015).

\bibitem{XiangY}Y. Xiang, F. X. Sun, M. Wang, Q. H. Gong, and Q. Y. He, Opt. Express \textbf{23}, 30104  (2015).

\bibitem{YangXH} X. H. Yang, Y. Ling, X. P. Shao, and M. Xiao, Phys. Rev. A \textbf{95}, 052303 (2017).

\bibitem{LiPRL2018}J. Li, S.-Y. Zhu, and G. S. Agarwal, Phys. Rev. Lett. \textbf{121}, 203601 (2018).

\bibitem{Palomaki}T. A. Palomaki, J. D. Teufel, R. W. Simmonds, K. W. Lehnert, Science \textbf{342}, 710 (2013).

\bibitem{Riedinger}R. Riedinger, A. Wallucks, I. Marinkovi\'{c}, C. L\"{o}schnauer, M. Aspelmeyer, S. K. Hong, and S. Gr\"{o}blacher, Nature (London) \textbf{556}, 473 (2018).

\bibitem{Korppi}C. F. Ockeloen-Korppi, E. Damsk\"{a}gg, J.-M. Pirkkalainen, M. Asjad, A. A. Clerk, F. Massel, M. J. Woolley, and M. A. Sillanp\"{a}\"{a}, Nature \textbf{556}, 478 (2018).

\bibitem{Massel1}F. Massel, T. T. Heikkil\"{a}, J.-M. Pirkkalainen, S. U. Cho, H. Saloniemi, P. J. Hakonen, and M. A. Sillanp\"{a}\"{a}, Nature (London) \textbf{480}, 351 (2011).

\bibitem{Bennett1}C. H. Bennett, G. Brassard, S. Popescu, B. Schumacher, J. A. Smolin, and W. K. Wootters, Phys. Rev. Lett. \textbf{76}, 722 (1996).

\bibitem{Bennett2}C. H. Bennett, D. P. DiVincenzo, J. A. Smolin, and W. K. Wootters, Phys. Rev. A \textbf{54}, 3824 (1996).

\bibitem{Teh}R. Y. Teh and M. D. Reid, Phys. Rev. A \textbf{90}, 062337 (2014).

\bibitem{RHcriterion} E. X. DeJesus and C. Kaufman, Phys. Rev. A \textbf{35}, 5288 (1987).

\bibitem{Gardiner}C. W. Gardiner and M. J. Collett, Phys. Rev. A \textbf{31}, 3761 (1985).

\bibitem{Tian2010}L. Tian and H. Wang, Phys. Rev. A \textbf{82}, 053806 (2010).

\bibitem{Tserkis2017}S. Tserkis and T. C. Ralph, Phys. Rev. A \textbf{96}, 062338 (2017).

\bibitem{Tserkis2019} S. Tserkis, S. Onoe, and T. C. Ralph, Phys. Rev. A \textbf{99}, 052337 (2019).

\bibitem{noteEOF} Note that the value of EOF is dimensionless, like any other entanglement measure or monotone, such as logarithmic negativity. As an entanglement measure, the higher the EOF between two continuous variable modes, the larger the entanglement is. For Gaussian states, the EOF could be related to two-mode squeezing between the continuous variable modes, and quantitatively, to the squeezing parameter $r_0$ given in Eq.~(\ref{eq:r0}). The larger $r_0$, the higher the EOF is.

\bibitem{Wallraff2013} C. Lang, C. Eichler, L. Steffen, J. M. Fink, M. J. Woolley, A. Blais, and A. Wallraff, Nat. Phys. \textbf{9}, 345 (2013).


\end{thebibliography}
\end{document}